\documentclass[manuscript]{aastex}

\shorttitle{Collapsed Cores in Globular Clusters}
\shortauthors{Djorgovski et al.}

\begin{document}


\title{The Stability of Core-Envelope Models for the Crab and the Vela pulsars}


\author{P. S. Negi}
\affil{Department of Physics, Kumaun  University,
    Nainital 263 002, India}

\email{psneginainital63@gmail.com; psnegi\_nainital@yahoo.com}





\begin{abstract}
The core-envelope analytic models of Negi et al. (1989) have been found to be consistent with both the mechanisms of glitch generation in pulsars, namely - (i) the starquake model and (ii) the vortex unpinning model. In Negi (2019), the author has been able to reproduce the observed values of glitch healing parameter, $ G_h (= I_{\rm core}/I_{\rm total}; G_h$ represents the fractional moment of inertia of the core component  in the starquake mechanism of glitch generation) for the Crab as well as for the Vela pulsar. In another study, by using the models discussed in  Negi et al. (1989), the author has obtained the minimum value of fractional moment of inertia of the crust about 7.4\% and larger for all the values of masses in the range - $1M_\odot - 1.96M_\odot$  considered for the Vela pulsar (Negi 2020a). The latter study of the author is found to be consistent with the recent requirement (on the basis of vortex unpinning model of the glitch generation) which refers:  The minimum fractional crustal moment of inertia of the Vela pulsar should be about 7\% for a  mass higher than about 1$M_\odot$. However, an important study which requires investigation of pulsational stability and gravitational binding of the models of Negi et al. (1989) has not been carried out so far. The present paper deals with such a study of the models (Negi et al. 1989) for all permissible values of the compactness parameter $u (\equiv M/a$, total mass to size ratio in geometrized units) and compressibility factor $Q$ (defined  in Tolman's  VII solution as: $ x = r^2/K^2 = r^2/a^2Q)$. It is seen that the configurations remain pulsationally stable and gravitationally bound for all permissible values of $u (\leq 0.25)$ and $Q$ ($0 < Q \leq 1.2$).
\end{abstract}

\keywords{Static Spherical Structures: Analytic Solutions -- Neutron Stars: Pulsars - Crab - Vela: Gravitational Binding, Pulsational Stability}

\section{Introduction}
\label{intro}
 Ever since the pioneering work of Oppenheimer and Volkoff (1939), the study regarding internal structure of Neutron Stars (NSs) has been a topic of vast astrophysical interest. Various models based on Equations of State (EOSs), derived from the knowledge of minimum physical constraints, are available in the literature (Rhoades and Ruffini 1974 , Brecher and Caporaso 1976, Hartle 1978). Since the details of the nuclear interactions are still  not well known beyond the density of $\sim 10^{14}\rm {g cm}^{-3}$, the NS models based on extrapolated EOSs around nuclear saturation density $E_n \sim 2.7 \times \rm {g cm}^{-3}$ and higher are widely discussed in the literature (Wiringa et al. 1988, Kalogera and Baym 1996, Haensel et al. 2006; and references therein). However, the importance and elegance of NS models based on analytic exact solutions of Einstein's field equations (Stephani et al. 2003) can't be denied. Among a variety of solutions for a static and spherically symmetric mass distribution, Tolman's VII solution is found to be the most realistic one (Durgapal et al. 1984, Lattimer and Prakash 2005). Negi and Durgapal (1999, 2001) have shown that the Tolman's VII solution with vanishing surface density remains stable and gravitationally bound for  $u > (1/3)$ and may be used to model NSs and pulsars. Later on, Moustakidis (2017) has verified the claim of  Negi and Durgapal (1999, 2001) regarding the stability of Tolman's VII solution with vanishing surface density.

Study of the phenomena of sudden spin-up rates in many pulsars (known as a glitch) is found to be  a powerful tool for understanding the internal structure of NSs. The most studied pulsars in this regard are the Crab and the Vela pulsars. At present, it is believed that the starquake mechanism of glitch generation (Baym and Pines 1971, Ruderman 1976, Alpar et al. 1996) is responsible for the Crab glitches (small size glitches, $\Delta \nu/\nu \sim 10^{-9}$) and another mechanism, known as vortex unpinning model (Anderson and Itoh 1975, Alpar et al. 1993) is appropriate for the glitch generation in the Vela pulsar (large size glitches, $\Delta \nu/\nu \sim 10^{-6}$).

The study of Negi et al. (1989) deals with the construction of  core-envelope models based on the  exact solutions of Einstein's field equations for static and spherical mass distribution. The core of the models is described by Tolman's VII solution, matched smoothly at the core-envelope boundary. The region of envelope is
described by Tolman's VI solution which is finally matched to vacuum Schwarzschild solution at the surface. The core-envelope boundary of the models is assured by matching of all the four variables - pressure ($P$), energy density ($E$) and both of the metric parameters $\nu$ and $\lambda$ without recourse to any computational method. The complete solutions for both the regions (the core and the envelope) are available in Negi et al. (1989). Recently, the author has shown that the models of Negi et al. (1989) can reproduce the observed values of glitch healing parameter for both the pulsars (that is, the Crab and the Vela pulsars) on the basis of the starquake model of glitch generation  (Negi 2019). This study of the author (Negi 2019) predicts the mass of the Crab pulsar in the range : 1.79$M_\odot -  1.88M_\odot$ whereas the mass of the Vela pulsar corresponds to a value $\sim$ 1.96$M_\odot$. In  another study based on the models of Negi et al. (1989), the author has been able to construct the models of the Vela pulsar which can reproduce the minimum fractional crustal moment of inertia about 7.4\% and larger for the mass range - 1$M_\odot - 1.96M_\odot$ considered for the Vela pulsar (Negi 2020a). It is to be pointed out here that the requirement of minimum fractional crustal moment of inertia about 7\% for a mass significantly higher than about 1$M_\odot$ for the Vela pulsar has recently been imposed due to crustal `entrainment' effect, discussed in the literature on the basis of the vortex unpinning explanation for the large Vela glitches (Andersson et al. 2012, Chamel 2013).

Therefore, it follows that the models of Negi et al. (1989) have been able to explain the observational data on the basis of the starquake as well as those on the basis of the vortex unpinning models for the Crab and the Vela pulsars (Negi 2019, Negi 2020a). However, the  gravitational binding and  pulsational stability of the models of  Negi et al. (1989) have not been discussed so far. Therefore, it would be interesting and also important to check whether the models discussed in Negi et al. (1989)  posses gravitationally bound structures and remain stable against small radial pulsations in the state of hydrostatic equilibrium ? Sec. 3 of the present paper is devoted to such a study. Results of this study are presented in Sec. 4. Sec. 5 summarizes the main findings of  this study.

\section{Gravitational Binding and Pulsational Stability of the Structures}
\label{sec:2}
We consider the metric corresponding to a static and spherically symmetric mass distribution in the following form
\begin{equation}
ds^2 = e^\nu dt^2 - e^\lambda dr^2 - r^2(d\theta^2 +{\rm sin}^2\theta d\phi^2)
\end{equation}
where $G = c = 1$ (that is, we are using geometrized units) and $\nu$ and $\lambda$ are functions of $r$ alone.
The relevant equations governing the core ($0\leq r \leq b$) and  the crust ($b\leq r \leq a$) are described respectively by Tolman's  VII and Tolman's  VI solutions of the Einstein's field equations for the metric (Eq.1) and are available in 
Negi et al. (1989). However, some relations relevant to the present study are defined as:

$u \equiv M/a$ is called the compactness parameter which is defined earlier; where
 the total mass is  given by
$M = \int_{0}^{a} 4\pi E r^2dr$; 
and $y = r/a$ is called the radial coordinate measured in units of configuration size.

$Q(\equiv K^2/a^2)$ is known as the compressibility factor which is also defined earlier; $K$ is a constant appearing in the Tolman's VII solution. The matching of various parameters at the core-envelope boundary yields (Negi et al. 1989)

$b^2/a^2Q = 5/6$, thus $(b/a)$ represents the boundary, $y_b$, of the core-envelope models and $b$ represents the core radius. The analytic  relation between central to boundary (energy) density, $E_0/E_b$, and central to surface (energy) density are given by:
$(E_0/E_b) = 6$; $(E_0/E_a) = 7.2/Q$.

The coefficient of gravitational binding $\alpha$ (that is, the gravitational binding energy per unit rest-mass) and the ratio of gravitational packing $\alpha_p$ ( that is, the ratio of gravitational energy to the total energy) can be obtained by using the equations (Zeldovich and Novikov 1978, Negi 2020b)
\begin{equation}
 \alpha = (M_r - M)/M_r = [(M_r/a) - (M/a)]/(M_r/a)
\end{equation}

\begin{equation}
\alpha_p = (M_p - M)/M = [(M_p/a) - (M/a)]/(M/a)
\end{equation}
where $M/a, M_r/a$ and $M_p/a$ are given by the relations
\begin{equation}
M/a = 4\pi \int_{0}^{1}Ea^2y^2dy
\end{equation}
\begin{equation}
M_r/a = 4\pi \int_{0}^{1}\rho a^2y^2 e^{\lambda/2}dy
\end{equation}
\begin{equation}
M_p/a = 4\pi \int_{0}^{1}Ea^2y^2 e^{\lambda/2}dy
\end{equation}
where $\rho = (P + E) e^{(\nu - \nu_a)/2}$ is called the rest-mass density (Durgapal and Pande 1980).The various quantities appearing in eqs. (2) - (6) may be calculated for various permissible values of $Q$ and $u$ by numerical integration from the centre of the configuration ($y = 0$) to the core-envelope boundary ($y_b = b/a$) by using Tolman's VII solution and from this boundary to the surface ($y = 1$) by employing Tolman's VI solution.

A core-envelope model may undergo a gravitational collapse for higher densities. To check whether the configuration is stable under small radial pulsations, a radial purturbation is introduced in the equilibrium  configuration which leads to radial oscillations in the structure.
The pulsational stability of the structures under such small radial perturbations can be judged by using the variational method (Chandrasekhar 1964, Negi 2020b; and references therein). For a stable configuration the pulsational frequency is given by
\begin{equation}
f = (1/2\pi)(A/B)^{1/2}
\end{equation}
where the functions $A$ and $B$ represent respectively, the potential energy and the kinetic energy with velocities replaced by displacements and are given by{\footnote 
{For simplification these expressions are obtained by using the `trial function' $\xi = re^{\nu/2}$, because this trial function is sufficient to judge the pulsational stability as obtained by using the trial function of the form of a power series ( Negi and Durgapal 1999, Negi 2007; and references therein) $\xi = b_1r(1 + a_1r^2 + a_2r^4 + a_3r^6)e^{\nu/2}$, where $a_1, a_2,$ and $a_3$ are arbitrary constants. Furthermore, the study of Knutsen (1989) also shows that the use of the trial function of the form of power series mentioned above (with suitable values of the arbitrary constants $a_1, a_2,$ and $a_3$ such that the appropriate boundary conditions may be satisfied) provide the results similar to those obtained by using the trial function $\xi = re^{\nu/2}$.}}
\begin{equation}
8\pi B/a^3 = \int_{0}^{1}(8\pi Pa^2 + 8\pi Ea^2)y^4e^{(3\lambda + \nu)/2}dy
\end{equation}
and
\begin{eqnarray}
8\pi A/a& = &\int_{0}^{1}y^2e^{3(\lambda + \nu)/2}\{ e^{-\lambda}[9(8\pi Pa^2 + 8\pi Ea^2)(dP/dE) \nonumber \\
& & + 4(8\pi a^2dP/dy)y - \frac{(8\pi a^2dP/dy)^2y^2}{(8\pi Pa^2 + 8\pi Ea^2)}] \nonumber \\
& & + 8\pi Pa^2(8\pi Pa^2 + 8\pi Ea^2)y^2 \}dy
\end{eqnarray}
Eqs. (8) and (9) may be computed by employing a fourth order Runge-Kutta method from the centre ($y = 0$) to the boundary ($y = b/a$) by using Tolman's VII solution and from the boundary ($y = b/a$) to the surface ($y = 1$) by using Tolman's VI solution which yield the values of function $(8\pi B/a^3)$ and $(8\pi a/A)$. On dividing values obtained by using eq.(9) by eq.(8), one gets the value of $a\omega$, where $\omega$ being the angular frequency of pulsation which follows from eq.(7). On computation, the positive values of pulsation frequencies would show that the average (constant) value of adiabatic index, $\gamma_{\rm ave}$, is larger than the minimum (critical) value of (constant) adiabatic index, $\gamma_{\rm crit}$, required for the stability of the structures (that is, $\gamma_{\rm ave} \geq \gamma_{\rm crit}$). Thus, we can safely conclude that the structures are stable under small radial perturbations. This is to be pointed out here that the use of the trial function $\xi = re^{\nu/2}$ in the above eqs. (8) and (9) safely assures the pulsational stability of the models considered in this study, because the present models correspond to the value of $u < 1/3$. For $u \geq 1/3$, the optimal trial function $\xi = re^{\nu/4}$ may be used for ascertaining the pulsational stability (see, e.g. Negi and Durgapal 2001, Moustakidis 2017, Negi 2020b) which is not required in the present study.
The various variables appear in eqs. (2) - (9) may be obtained from Negi et al. (1989) in the following form

\subsection{The Core: $0 \leq y \leq (b/a)$}
\label{sec:1}
\begin{equation}
8\pi Ea^2 = 8\pi E_0a^2(1 - x)
\end{equation}
where $x = y^2/Q$; and $8\pi E_0a^2 = 72u/5Q$
\begin{equation}
e^{-\lambda} = 1 - (24u/25)(5 - 3x)x
\end{equation}
\begin{equation}
e^{\nu/2} = C_1{\rm cos}(w/2) + C_2{\rm sin}(w/2)
\end{equation}
\begin{equation}
8\pi Pa^2 = Ma^2\frac{C_2{\rm cos}(w/2) - C_1{\rm sin}(w/2)}{ C_1{\rm cos}(w/2) + C_2{\rm sin}(w/2)} - Na^2
\end{equation}
\begin{equation}
 Ma^2 = 12(2u)^{1/2}e^{-\lambda/2}/5Q
\end{equation}
\begin{equation}
 Na^2 = (24/25Q)u(5 - 3x)
\end{equation}
\begin{eqnarray}
w& = & \rm{ln} \{ x - (5/6) \nonumber \\
& & + \left [ ( 25/72u) - (1/3)(5 - 3x)x \right ]^{1/2} \}
\end{eqnarray}
\begin{equation}
-8\pi a^2 \frac{dE}{dy} = 8\pi E_0a^2(2y/Q)
\end{equation}

\begin{equation}
-8\pi a^2 \frac{dP}{dy} =(1/2)( 8\pi Pa^2 + 8\pi Ea^2)(8\pi Pa^2 + Na^2)e^\lambda y
\end{equation}
\begin{equation}
8\pi \rho a^2 = (8\pi Pa^2 + 8\pi Ea^2)e^{\nu/2}(1 - 2u)^{-1/2}
\end{equation}

\subsection{The Envelope: $(b/a) \leq y \leq 1$}
\label{sec:2}
\begin{equation}
8\pi Ea^2 = 2u/y^2
\end{equation}
\begin{equation}
e^{-\lambda} = (1 - 2u)
\end{equation}
\begin{equation}
e^{\nu/2} = \frac {y^{(1 - n)}[(1 + n)^2 - (1 - n)^2y^{2n}]}{4n(2 - n^2)^{1/2}}
\end{equation}
\begin{equation}
8\pi Pa^2  = \frac{2u(1 - n^2)(1 - y^{2n})}{y^2[(1 + n)^2 - (1 - n)^2y^{2n}]}
\end{equation}

\begin{equation}
-8\pi a^2 \frac{dE}{dy} = 4u/y^3
\end{equation}

\begin{equation}
-8\pi a^2 \frac{dP}{dy} =(1/2)( 8\pi Pa^2 + 8\pi Ea^2)(8\pi Pa^2y^2 + 1 - e^{-\lambda})e^\lambda y^{-1}
\end{equation}
\begin{equation}
8\pi \rho a^2 = (8\pi Pa^2 + 8\pi Ea^2)e^{\nu/2}(1 - 2u)^{-1/2}
\end{equation}
where $n^2 = (1 - 4u)/(1 - 2u)$; and the constants $C_1$ and $C_2$ are given [1]
\begin{equation}
C_1 =  A_1{\rm cos}(w_b/2) -  B_1{\rm sin}(w_b/2)
\end{equation}
\begin{equation}
C_2 =  A_1{\rm sin}(w_b/2) +  B_1{\rm cos}(w_b/2)
\end{equation}
where
\begin{equation}
w_b =  \rm{ln}(5/6)[(1/2u) - 1]^{1/2} 
\end{equation}
\begin{equation}
A_1 =   \frac {y_b^{(1 - n)}[(1 + n)^2 - (1 - n)^2y_b^{2n}]}{4n(2 - n^2)^{1/2}}
\end{equation}
\begin{equation}
B_1 =  (A_1/M_bb^2) \left ( \frac{2u( 1 - n^2)(1 - y_b^{2n})} {(1 + n)^2 - (1 - n)^2y_b^{2n}} + N_bb^2 \right )
\end{equation}
and
\begin{equation}
M_bb^2  = 2(2u)^{1/2}(1 - 2u)^{1/2}
\end{equation}
\begin{equation}
N_bb^2 =  2u
\end{equation}
\section{Results and Conclusions}
\label{sec:4}
Figures 1 - 4 give the plots between $a\omega$ and $Q$ for $u$ values  0.10,\, 0.15,\, 0.20 and 0.24999 respectively. The positive values of $a\omega$ indicate that the structures are pulsationally stable for all values of $u$ permissible for the models considered in the study of Negi et al.(1989).
The binding energy coefficients, $\alpha$ and $\alpha_p$, of the models considered in the present study are shown is Figures 5 - 8 for $u$ values in the range  0.10 - 0.24999. The positive values of $\alpha$ (since $\alpha_p$ always remains positive) indicate that the structures are gravitationally bound for all possible values of $Q$ and $u$. As $Q \rightarrow 1.2$ the values of $\alpha$ and $\alpha_p$ become closer to each other. However, this value of $Q$ corresponds to a slower variation of density inside the structure (which corresponds to a structure with a negligible envelope, i.e. the entire configuration is represented by Tolman's VII solution) so much so that the rest- mass density and the energy density become almost equal. Furthermore, it may be noted that $\alpha$ is continuously increasing with $Q$ for all the values of $u$ considered in the range 0.10 - 0.24999 which means that the structures are also pulsationally stable together with the fact that they are also gravitationally bound which is the outcome of binding energy criterion of fluid stars which states that the configurations remain pulsationally stable upto the first maxima in the binding energy curve (Zeldovich and Novikov 1978, Shapiro and Teukolsky 1983).

\section{Summary}
\label{sec:5}
A core-envelope massive configuration corresponding to a core described by Tolman's VII solution and the envelope is given by Tolman's VI solution has been investigated under small radial pulsations and the binding-energy coefficients ($\alpha$ and $\alpha_p$) have been calculated for various allowed values of compactness parameters, $u$ and the compressibility factor $Q$. The study shows that the models of Negi et al.(1989) for which all the four variables $P, E , \nu$ and $\lambda$ along with ($\rm d\nu/\rm dr$) and ($\rm d\lambda/\rm dr$) are continuous at the core-envelope boundary $r = b$ represent
 causally consistent, gravitationally bound and pulsationally stable structures.

Thus, these gravitationally bound and pulsationally stable models are capable of explaining the glitch healing parameters on the basis of starquake model of glitch generation for both the pulsars, namely the Crab pulsar and the Vela pulsar and predict the mass for the Crab pulsar about $1.8M_\odot$ and for the Vela pulsar about $1.96M_\odot$ for the surface density $2\times 10^{14}\rm {g cm}^{-3}$ (like, Brecher and Caporaso 1978) considered in the study of Negi (2019). Furthermore, on the basis of vortex unpinning model of glitch generation, the models considered in this study are capable of providing the recent required  minimum value of  fractional crustal moment of inertia about 7\% and higher for all values of the masses in the range $1M_\odot - 1.96M_\odot$ assigned for the Vela pulsar(Negi 2020a).


\section {Conflict of Interest Statement:} The Author declares that there is no conflict of interest.

\begin{figure*}
\caption{Variation of $a\omega$ with $Q$ for $u$ value 0.10.}
\includegraphics[width=0.75\textwidth]{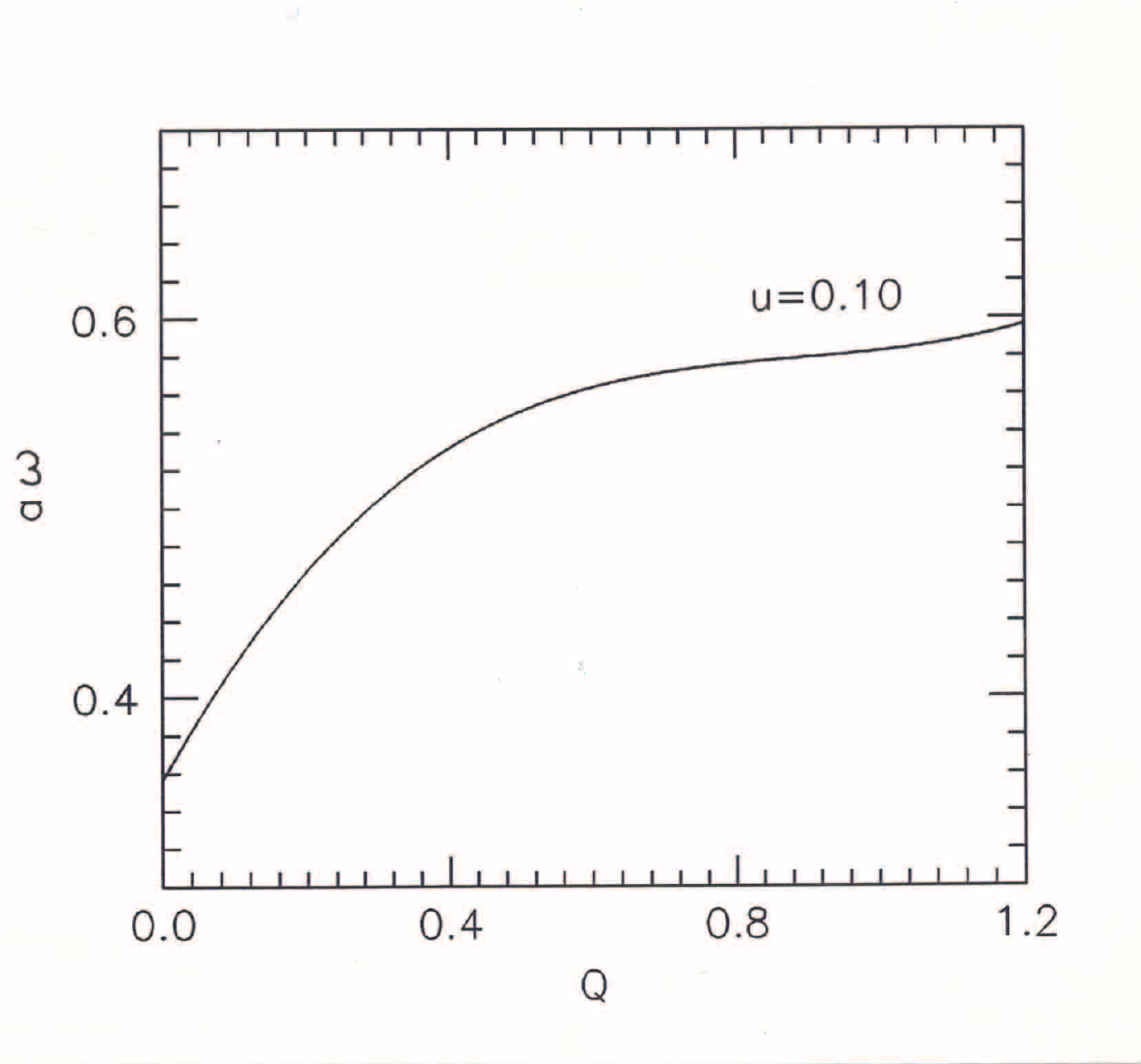}
\label{fig:1} 
\end{figure*}

\begin{figure*}
\caption{Variation of $a\omega$ with $Q$ for $u$ value 0.15.}
\includegraphics[width=0.75\textwidth]{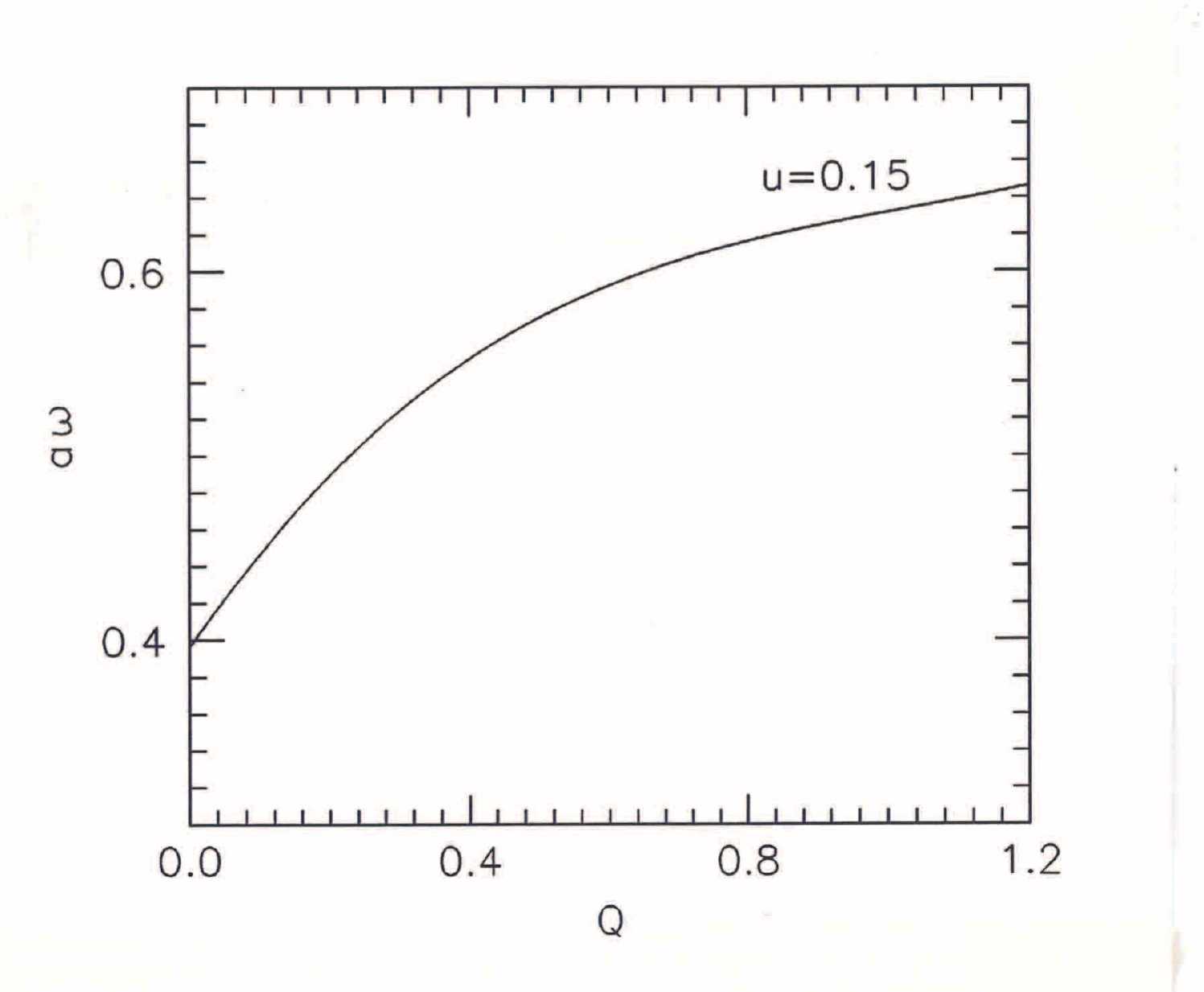}
\label{fig:2} 
\end{figure*}

\begin{figure*}
\caption{Variation of $a\omega$ with $Q$ for $u$ value 0.20.}
\includegraphics[width=0.75\textwidth]{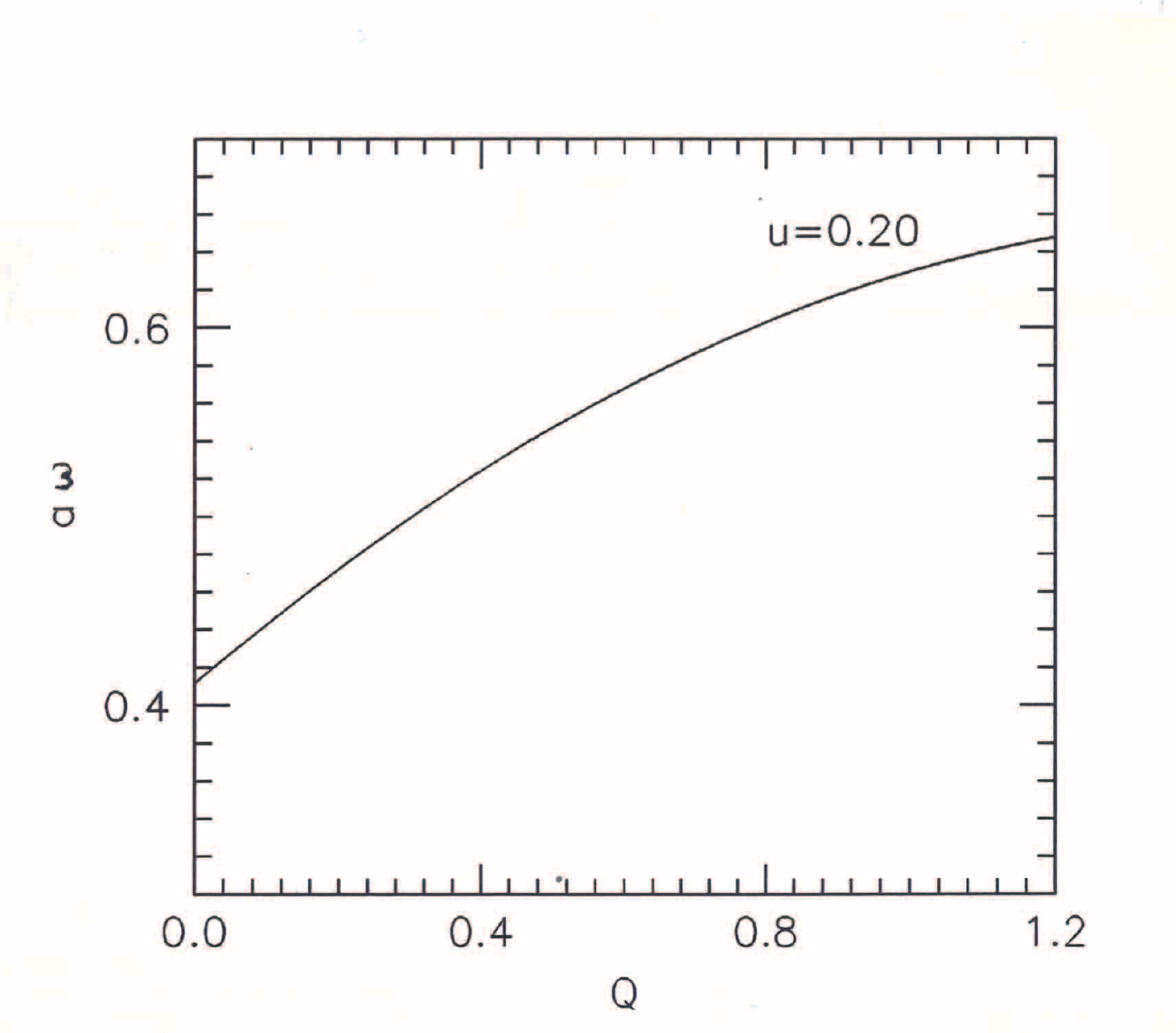}
\label{fig:3} 
\end{figure*}

\begin{figure*}
\caption{Variation of $a\omega$ with $Q$ for $u$ value 0.24999.}
\includegraphics[width=0.75\textwidth]{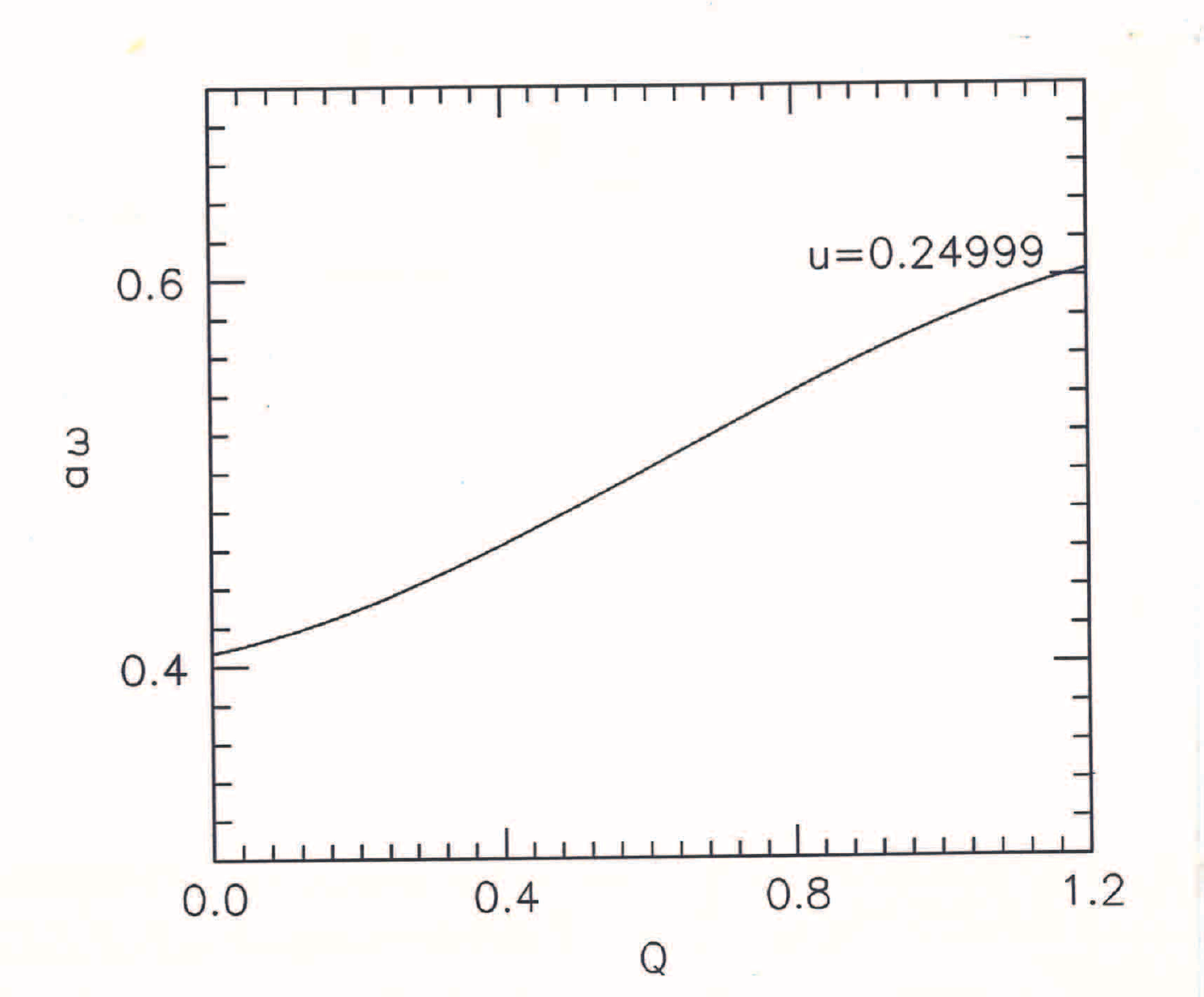}
\label{fig:4} 
\end{figure*}
\begin{figure*}
\caption{Variation of $\alpha, \alpha_p$ with $Q$ for the value of $u = 0.10$.}
\includegraphics[width=0.75\textwidth]{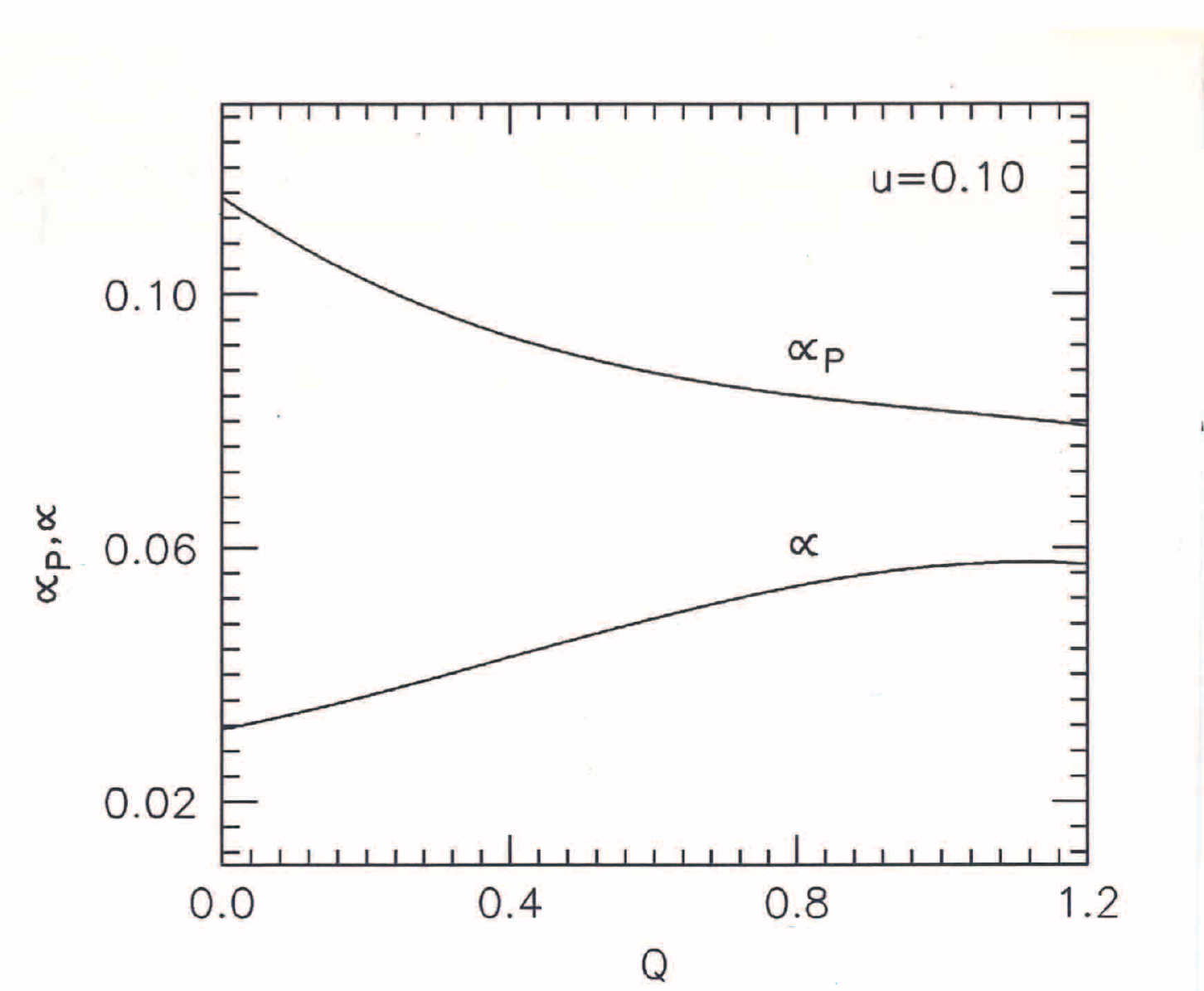}
\label{fig:5} 
\end{figure*}

\begin{figure*}
\caption{Variation of $\alpha, \alpha_p$ with $Q$ for the value of $u = 0.15$.}
\includegraphics[width=0.75\textwidth]{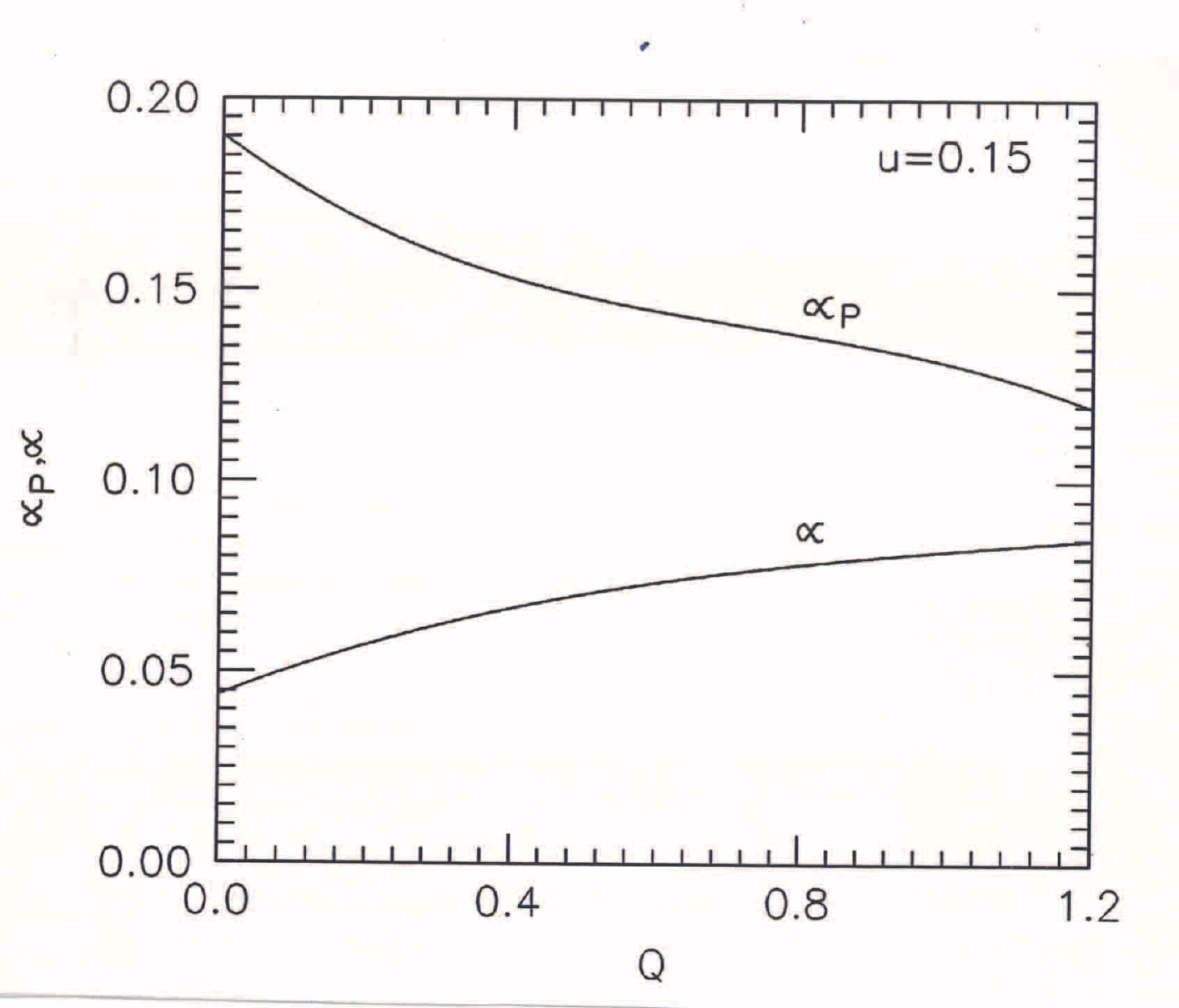}
\label{fig:6} 
\end{figure*}

\begin{figure*}
\caption{Variation of $\alpha, \alpha_p$ with $Q$ for the value of $u = 0.20$.}
\includegraphics[width=0.75\textwidth]{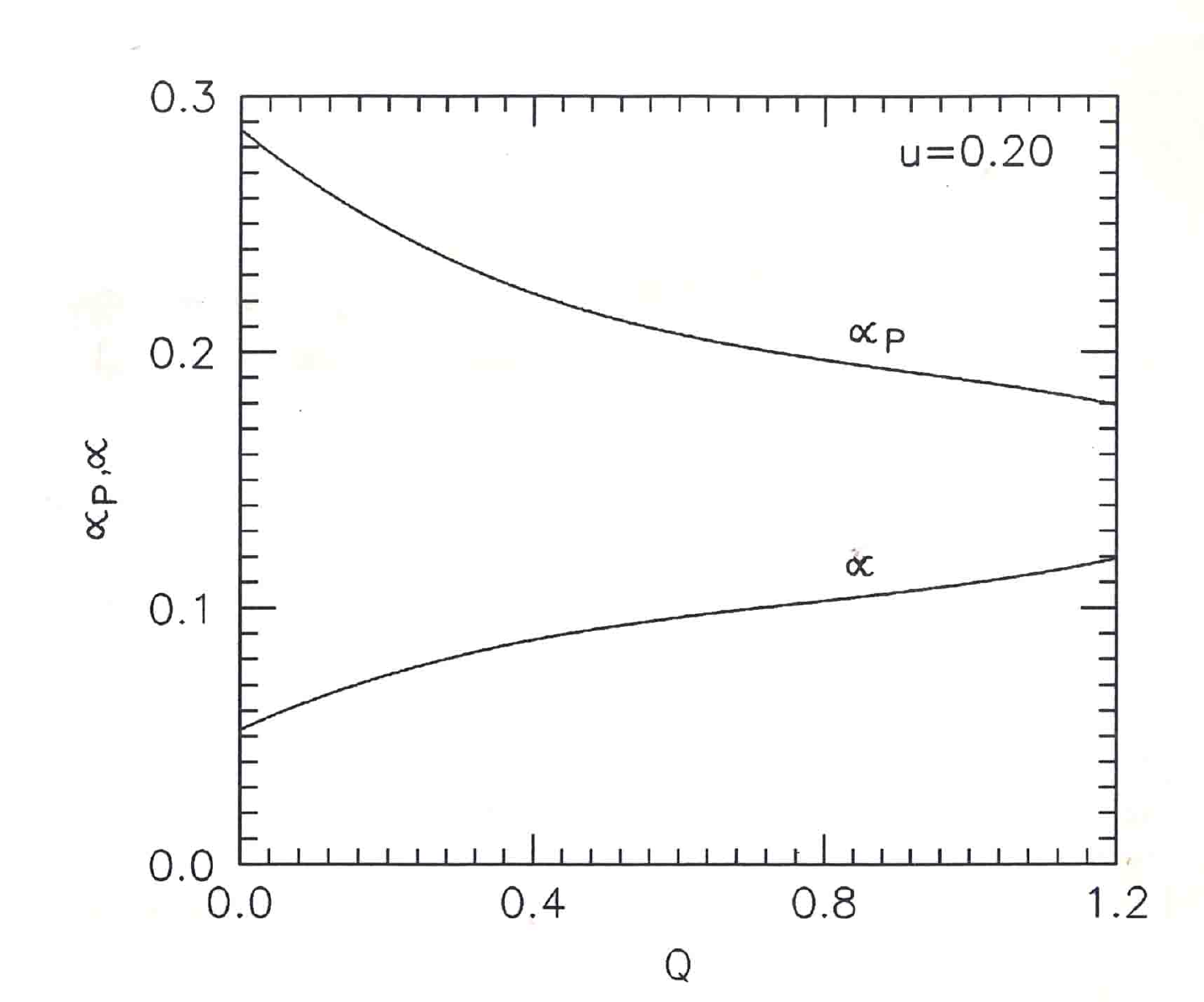}
\label{fig:7} 
\end{figure*}
\begin{figure*}
\caption{Variation of $\alpha, \alpha_p$ with $Q$ for the value of $u = 0.24999$.}
\includegraphics[width=0.75\textwidth]{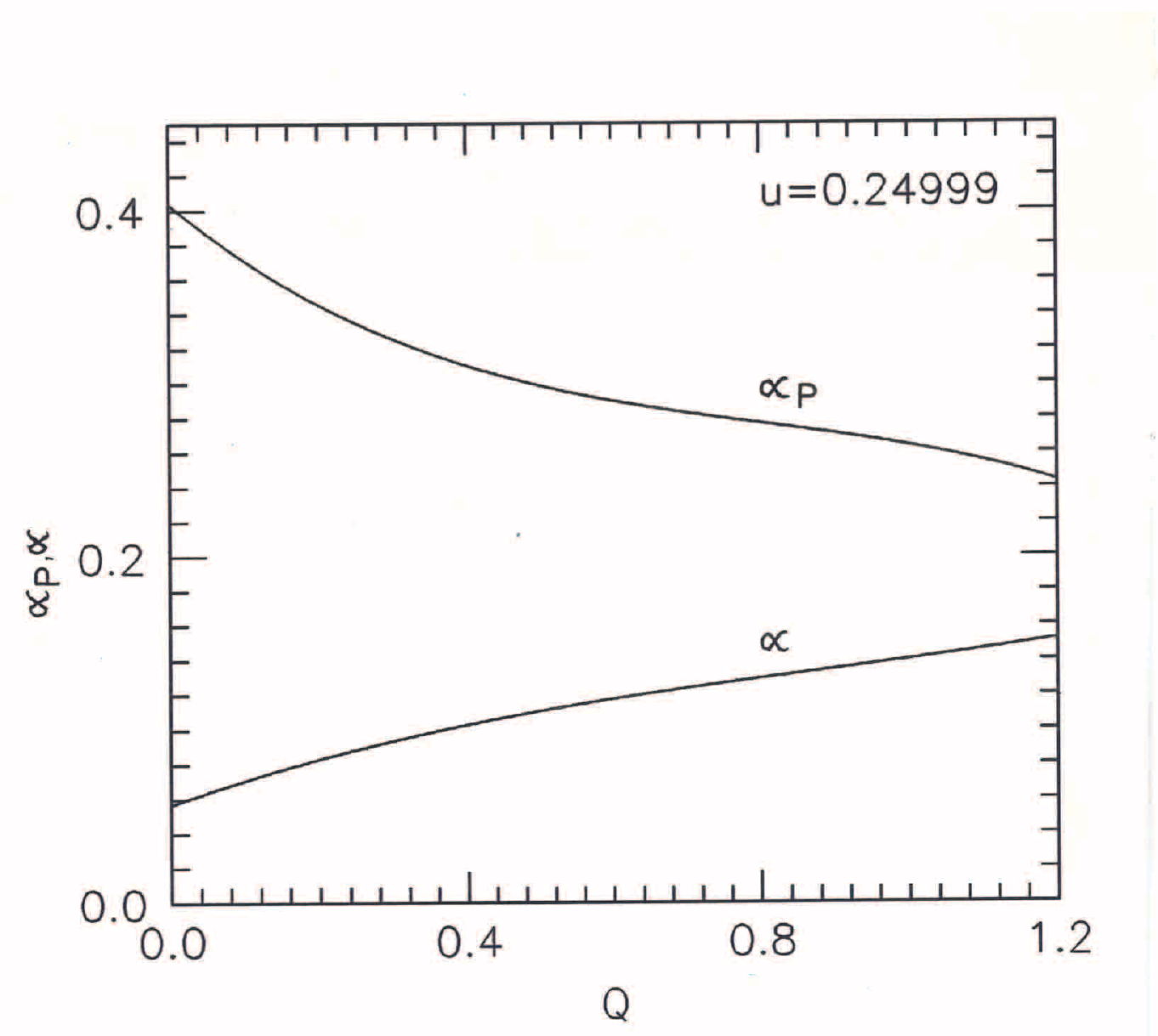}
\label{fig:8} 
\end{figure*}

\end{document}